\documentclass[aps,twocolumn,superscriptaddress,showpacs]{revtex4-1} %prl,twocolumn
\usepackage{graphicx}
\usepackage{graphicx,epstopdf,color}
\usepackage{amsfonts}
\usepackage{amsmath,amssymb,mathrsfs}
\usepackage{bm}
\usepackage{pst-grad}

\usepackage{dcolumn}

\def \ba122{BaFe$_2$As$_2$}
\def \basp{BaFe$_2$(As$_{1-x}$P$_{x}$)$_2$}
\def \sr327{Sr$_3$Ru$_2$O$_7$}

\begin{document}

\title{Magnetoresistance near a quantum critical point}

\author{I. M. Hayes}
\affiliation{Department of Physics, University of California, Berkeley, California 94720, USA}

\author{Nicholas P. Breznay}
\affiliation{Department of Physics, University of California, Berkeley, California 94720, USA}
\affiliation{Materials Science Division, Lawrence Berkeley National Laboratory, Berkeley, California 94720, USA}

\author{Toni Helm}
\affiliation{Department of Physics, University of California, Berkeley, California 94720, USA}
\affiliation{Materials Science Division, Lawrence Berkeley National Laboratory, Berkeley, California 94720, USA}

\author{Philip Moll}
\affiliation{Department of Physics, University of California, Berkeley, California 94720, USA}
\affiliation{Materials Science Division, Lawrence Berkeley National Laboratory, Berkeley, California 94720, USA}

\author{Mark Wartenbe}
\affiliation{National High Magnetic Field Laboratory, Florida State University, Tallahassee, FL 32310, USA}

\author{Ross D. McDonald}
\affiliation{Los Alamos National Laboratory, Los Alamos, NM 87545, USA}

\author{Arkady Shekhter}
\affiliation{National High Magnetic Field Laboratory, Florida State University, Tallahassee, FL 32310, USA}

\author{James G. Analytis}
\affiliation{Department of Physics, University of California, Berkeley, California 94720, USA}
\affiliation{Materials Science Division, Lawrence Berkeley National Laboratory, Berkeley, California 94720, USA}

\begin{abstract}
 In metals near a quantum critical point, the electrical resistance is thought to be determined by the lifetime of the carriers of current, rather than the scattering from defects. The observation of $T$-linear resistivity suggests that the lifetime only depends on temperature, implying the vanishing of an intrinsic energy scale and the presence of a quantum critical point. Our data suggest that this concept extends to the magnetic field dependence of the resistivity in the unconventional superconductor BaFe$_2$(As$_{1-x}$P$_{x}$)$_2$ near its quantum critical point. We find that the lifetime depends on magnetic field in the same way as it depends on temperature, scaled by the ratio of two fundamental constants $\mu_B/k_B$. These measurements imply that high magnetic fields probe the same quantum dynamics that give rise to the $T$-linear resistivity, revealing a novel kind of magnetoresistance that does not depend on details of the Fermi surface, but rather on the balance of thermal and magnetic energy scales. This opens new opportunities for the investigation of transport near a quantum critical point by using magnetic fields to couple selectively to charge, spin and spatial anisotropies.
\end{abstract}

\maketitle

The conventional description of metals depends on the existence of a robust Fermi surface in momentum space, a locus of free-electron-like quasiparticles that determine most properties of a metal, including the ability to pass electric current\cite{landau_FL_1957}. In this picture, the temperature dependence of the quasiparticle scattering rate ($\sim T^2/E_F$ where $E_F$ is the Fermi energy) arises from rare quasiparticle decay events and provides a weak addition to the elastic scattering. However, many correlated electron metals, particularly those thought to be near a quantum critical point (QCP), display a striking $T$-linear dependence of the electrical resistivity \cite{cooper_anomalous_2009, borzi_formation_2007, kasahara_evolution_2010, custers_break-up_2003, doiron-leyraud_fermi-liquid_2003}. Assuming that it is the scattering rate that is responsible for all the temperature dependence of resistivity, these experimental observations suggest a linear temperature dependence of the quasiparticle scattering rate, $\hbar/\tau = \alpha k_BT$, with $\alpha$ close to unity across a broad range of materials \cite{bruin_similarity_2013}. Although the physical mechanism behind this is not fully understood \cite{dalidovich_nonlinear_2004,green_nonlinear_2005}, it is argued that because temperature is the dominant energy scale in the system, $T$ sets the scattering rate near a QCP, resulting in the observed $T$-linear resistivity \cite{varma_phenomenology_1989, sachdev_universal_1992, varma_pseudogap_1999, sachdev_fluctuating_2009, davison_holographic_2014}.  
Since this argument only depends on temperature's being the dominant energy scale, it raises the question of whether the same intuition can be extended to magneto-transport phenomena in quantum critical systems, with the magnetic field playing the role of temperature.

In conventional metals the effect of a magnetic field on the quasiparticle scattering rate is negligible; the magnetoresistance arises from quasiparticles traversing the Fermi surface in cyclotron orbits under the influence of the Lorentz force. The field dependence of the resistance (MR) therefore is a sensitive function of the morphology of the FS, the curvature of which causes a suppression of the average velocity in the direction of the current \cite{pippard_magnetoresistance_2009}. The scattering time only affects the MR by limiting the path length of a cyclotron orbit, which in turn leads to Kohler's rule scaling~\cite{pippard_magnetoresistance_2009}. Even in models of MR near a QCP, the morphology of the Fermi surface generally plays a crucial role \cite{rosch_magnetotransport_2000, schofield_quasilinear_2000}. In contrast, we hypothesize that in quantum critical metals the magnetic field can set the energy scale of the scattering rate in the same way as temperature, such that $\hbar/\tau=\eta\mu_BB$. This would result in $B$-linear MR and could provide a natural explanation for the $B$-linear MR that has been observed in many correlated metals thought to be near a QCP \cite{weickert_low-temperature_2006, rourke_phase-fluctuating_2011,krusin-elbaum_interlayer_2010, butch_quantum_2012}.

\begin{figure}[ht]
\includegraphics[width=7cm]{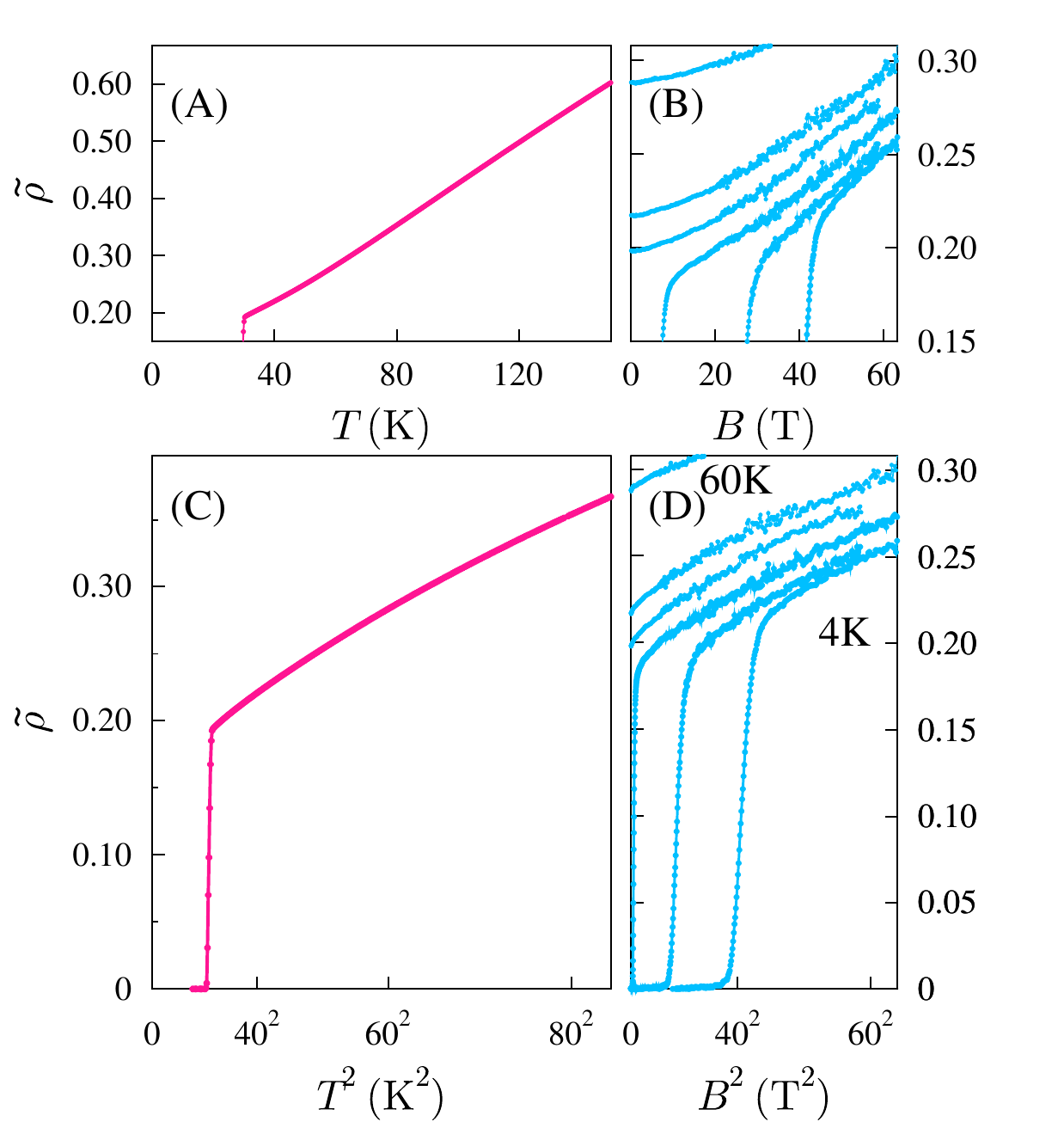} 
\caption{{\bf Similarity of the functional form of $\tilde{\rho}(T)$ and $\tilde{\rho}({\bf B})$ in \basp.} {\bf A.} The temperature dependence of the normalized resistivity $\tilde{\rho}(T)=\rho(T)/\rho({273K})$ is $T$-linear at high-$T$, which appears as sub-linear behavior in {\bf C} when plotted against $T^2$. {\bf B.} The MR qualitatively looks $B$-linear at high-$B$, particularly at low temperatures. The similarity with the resistance as a function of temperature is striking when plotted against $B^2$, as can be seen by comparing {\bf C \& D.} The MR curves were taken at temperatures 4K, 14K, 25K, 31K, 38K and 60K.}
\label{fig:TsqBsq} 
\end{figure}

The unconventional superconductor \basp\, is a system that is known to show $T$-linear resistivity in a broad range of temperature and composition $x$ \cite{analytis_transport_2014}. The parent compound, BaFe$_2$As$_2$, is a tetragonal paramagnet at high temperatures, transitioning to an orthorhombic antiferromagnet at around 140K\cite{rotter_spin-density-wave_2008}. With the substitution of  P for As the magnetic transition temperature is suppressed, heading towards zero temperature at around $x \approx 0.3$ (the doping at which there is thought to be a quantum critical point \cite{dai_iron_2009, kasahara_evolution_2010, nakai_p31_2010, shishido_evolution_2010}), where the superconducting $T_c$ reaches a  maximum of $\sim$30~K. These materials are perhaps the cleanest of the doped Fe-pnictide systems, showing quantum oscillations at compositions $x>0.4$ \cite{shishido_evolution_2010}. They are therefore an ideal material family in which to study magnetoresistance of a metal near a quantum critical point.

\begin{figure*}[ht]
\includegraphics[width=16cm]{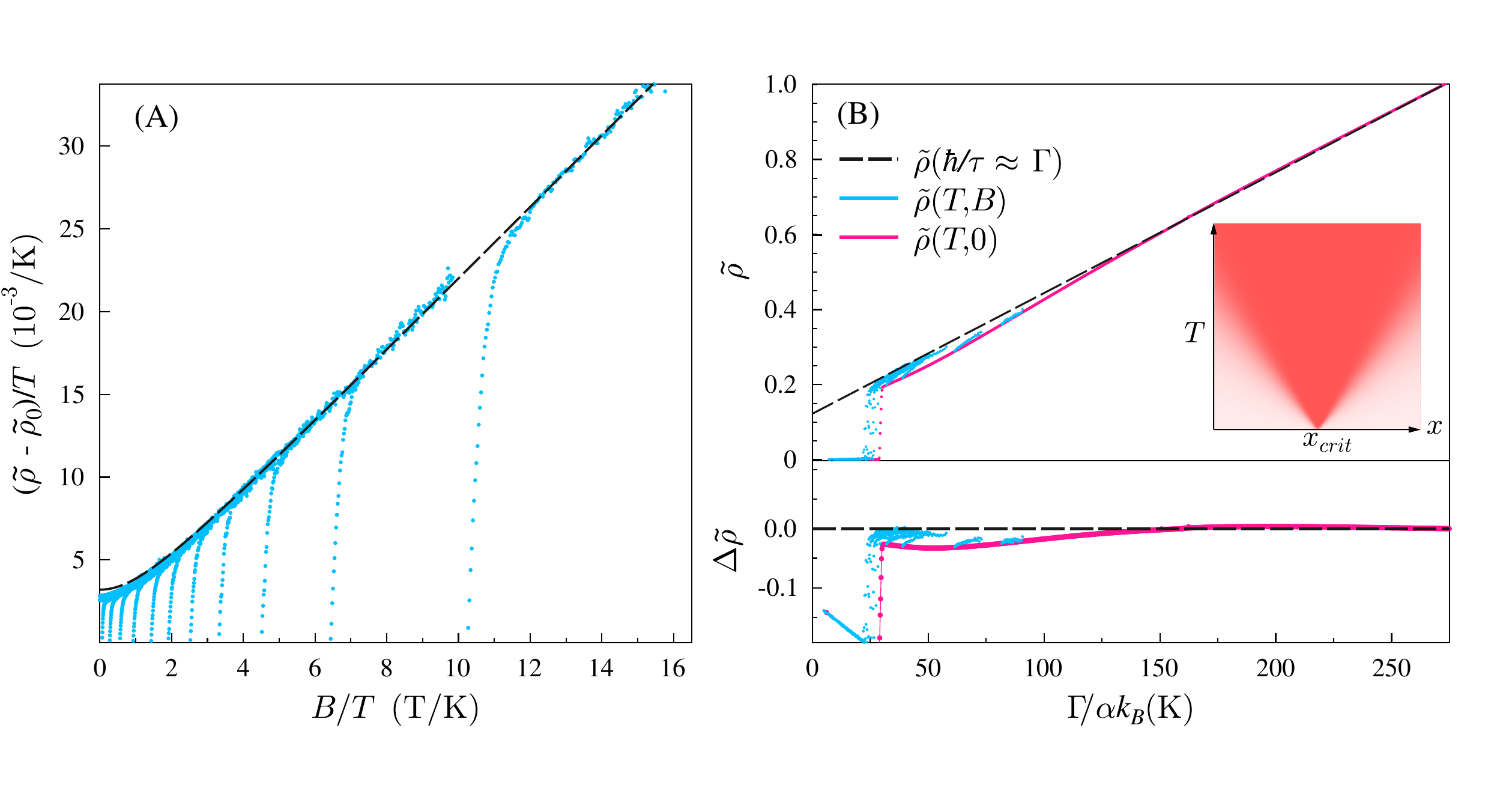} 
\caption{{\bf $B - T$ scaling in \basp\,at $x=0.31$.} {\bf A.} After subtracting the residual resistivity $\tilde{\rho}_0$, we plot the MR divided by temperature versus $B/T$. The data are well approximated by a function of the form $\propto\sqrt{1 + c(B/T)^2}$, with some numeric parameter $c$, plotted as the grey dashed line.  {\bf B.}  The MR (blue) and zero field resistance (pink) plotted as a function of $\Gamma$ up to room temperature. The extrapolation of the high-temperature, linear region is shown as a grey, dashed line. The lower panel the MR with this linear component subtracted, showing that the MR asymptotically approaches the line $\hbar/\tau = \alpha k_B T$. The inset shows a schematic quantum critical phase diagram with the region of $T$ - linear resistivity marked in red. Note that the data shown here include that shown in Fig. \ref{fig:TsqBsq}.}
\label{fig:Scaling} 
\end{figure*}

We have measured the in-plane MR of \basp\, up to 65 Tesla, with the field applied parallel to the crystallographic $c$-axis. Figure \ref{fig:TsqBsq} (A) shows the normalized resistivity $\tilde{\rho}=\rho(T)/\rho({273K})$ as a function of temperature for a sample (x = 0.31) that is near the putative critical point and is strongly T-linear at high temperature. The small amount of curvature at lower temperatures should not affect our analysis since we are interested in high field and high temperature effects.\cite{analytis_transport_2014}. Figure \ref{fig:TsqBsq} (B) shows the MR, which exhibits an analogous high field $B$-linear dependence. The similarity between the transport as a function of $T$ and $B$ is even more striking when the data is plotted as $T^2$ and $B^2$ on similar vertical scales, as shown in Fig. \ref{fig:TsqBsq} (C) and (D). Ideally we would measure the MR of the normal state over a wide range of magnetic field at zero (or at least very low) temperature but superconductivity cuts off the data below the upper critical field $\mu_0H_{c2}$, restricting us to a narrow range of the high magnetic field region where $B$-squared and $B$-linear behaviors are difficult to distinguish. However, we observe that the MR qualitatively appears more $B$-linear at low temperature and more $B$-squared at high temperature (a fact that is widely known in these and other unconventional superconductors \cite{doiron_leyraud_correlation_2009, rourke_phase-fluctuating_2011, rullier-albenque_high-field_2011}). This gives us an important clue as to how magnetic field and temperature influence transport - they appear to compete to set the scale of the scattering.

To analyze the form of this competition, we look for scaling behavior of the MR as a function of $B$ and $T$ by plotting $\tilde{\rho}/T$ versus $B/T$. Note that we remove the residual resistivity $\tilde{\rho}_0$ so that only the temperature dependent part of the resistivity enters the analysis (see Supplementary Information for details on the determination of $\tilde{\rho}_0$). When plotted this way, the data appear to collapse to a single curve that is well described by a hyperbolic function of $B/T$, as shown in Fig. \ref{fig:Scaling} A. We therefore formulate the following ansatz
\begin{equation}
{\hbar\over\tau}  = \sqrt{(\alpha k_BT)^2+\left(\eta{\mu_B} B \right)^2}\equiv{\Gamma},
\label{eq:phenomenon}
\end{equation}
where $\alpha$ and $\eta$ relate the scattering rate directly to the temperature and magnetic field scales respectively \cite{schroder_onset_2000}. A similar ansatz has been used to describe the temperature dependence of the optical conductivity in  URu$_2$Si$_2$ \cite{nagel_optical_2012}, the magnetic susceptibility of CeCu$_{6-x}$Au$_x$ \cite{schroder_onset_2000} and the magnetic fluctuations of La$_{1.86}$Sr$_{0.14}$CuO$_4$ \cite{aeppli_nearly_1997}. Furthermore, this expression naturally captures the limiting cases of high field and high temperature. As long as $T\gg B$ the scattering rate will be $T$-linear, and vice versa. Between these two limits magnetic field and temperature compete to set the scale of the scattering.

\begin{figure*}[ht]
\includegraphics[width=18cm]{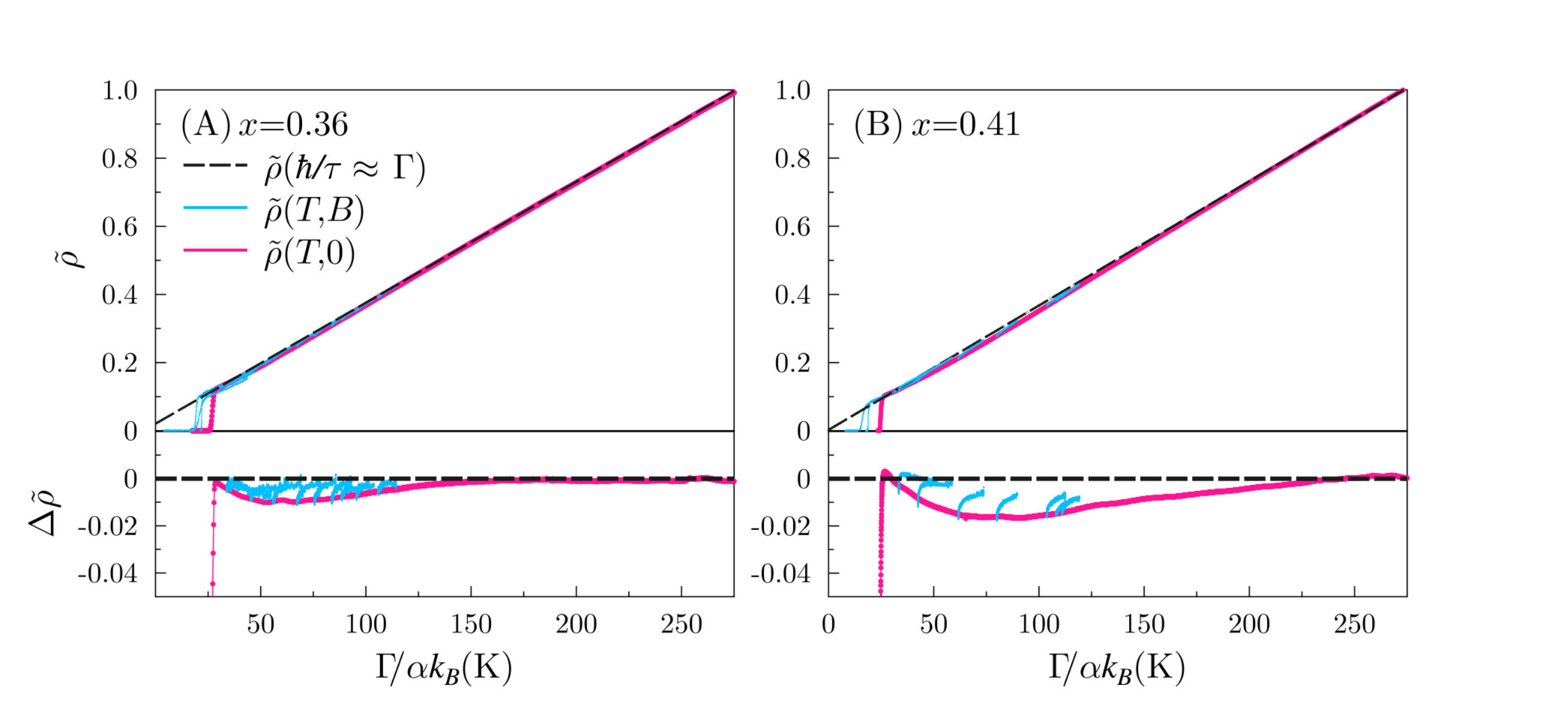} 
\caption{{\bf Magnetoresistance of \basp\, at two compositions near optimal $T_c$.} A figure analogous to Fig. \ref{fig:Scaling} (B) showing the resistance as a function of $\Gamma$ (using
$\eta = \alpha$) at {\bf A.} $x = 0.36$, and {\bf B.} $x = 0.41$. Lower panels show the same data with the linear component, $\tilde{\rho} \propto \Gamma$ (black dashed line), subtracted, and with the lowest temperature curves removed for clarity.}
\label{fig:rvg2} 
\end{figure*}

In order to understand the degree to which magnetic field and temperature each influence the scattering, it is necessary to determine the scale factor, $\eta / \alpha$. The variable $\alpha$ can be determined by the slope of the resistance versus temperature at zero field. However, because there is a cross over from $T$-linear to $T$-squared behavior as temperature is lowered, we use the high-$T$ limit of the resistivity data to determine $\alpha$. Similarly, we look at the high-$B$ limit to determine $\eta$, choosing the lowest temperature curve so that magnetic field is certain to be the dominant energy scale. Comparing the slope of resistance versus $T$ and $B$ in these regions, we find that $\eta/\alpha = 1.01\pm 0.07$ (see Supplementary Information for details). With the value of the scale factor $\eta/\alpha$ determined, we can plot all of the MR data, together with the zero field resistance, as a function of $\Gamma$ (Fig. \ref{fig:Scaling} (B)). We observe that for any combination of field and temperature, the resistance always approaches the same $\Gamma$-linear dependence, as described by Eq. \ref{eq:phenomenon}. Remarkably, this behavior holds for a range of dopings near the critical point (see Fig. \ref{fig:rvg2}). This is not expected in a conventional metal, where the magnitude and form as a function of $T$ and $B$ will in general be unrelated, each depending on different details of the Fermi surface.

\begin{figure*}[ht]
\includegraphics[width=18cm]{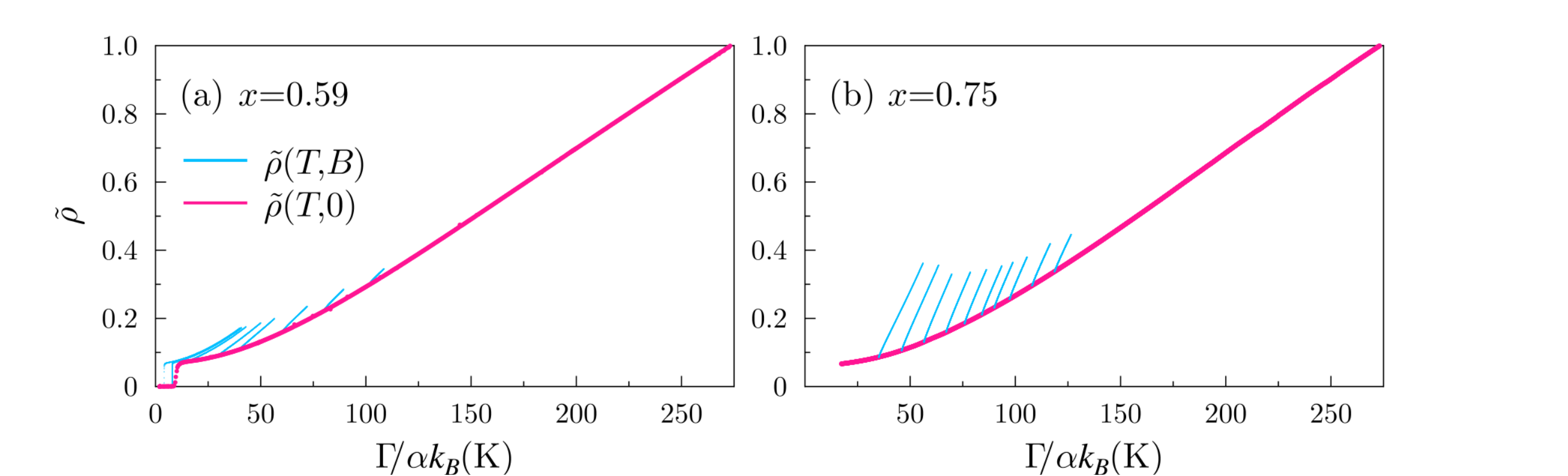} 
\caption{{\bf Magnetoresistance of \basp\, at two values of $x$, far from optimal doping.} A figure analogous to Fig. \ref{fig:Scaling} (B) showing the resistance as a function of $\Gamma$ with $\eta/\alpha=1$ at {\bf A.} $x = 0.59$ and {\bf B.} $x = 0.75$. Our analysis suggests that no choice of $\eta/\alpha$ will cause the data to collapse onto a single function of $\Gamma$. These samples are far from optimal doping and have low superconducting transition temperatures, a region of the \basp\,phase diagram which is known to be Fermi liquid-like~\cite{kasahara_evolution_2010}.}
\label{fig:overdoped_rvg} 
\end{figure*}

At temperatures below $T_c$, we note a deviation from Eq. \ref{eq:phenomenon}. Although we cannot account for this in our ansatz, it seems likely that the deviation is due to the presence of other energy scales related to superconductivity. At compositions far from the QCP, we expect this behavior to break down completely as the materials crossover to conventional Fermi liquids\cite{kasahara_evolution_2010}. This is indeed observed, as shown in Fig. \ref{fig:overdoped_rvg} for compositions well beyond $x_c$. A more detailed description of the relevance of Eq. \ref{eq:phenomenon} as a function of $x$ will require a much more extensive study.

\begin{figure}[ht]
\includegraphics[width=7cm]{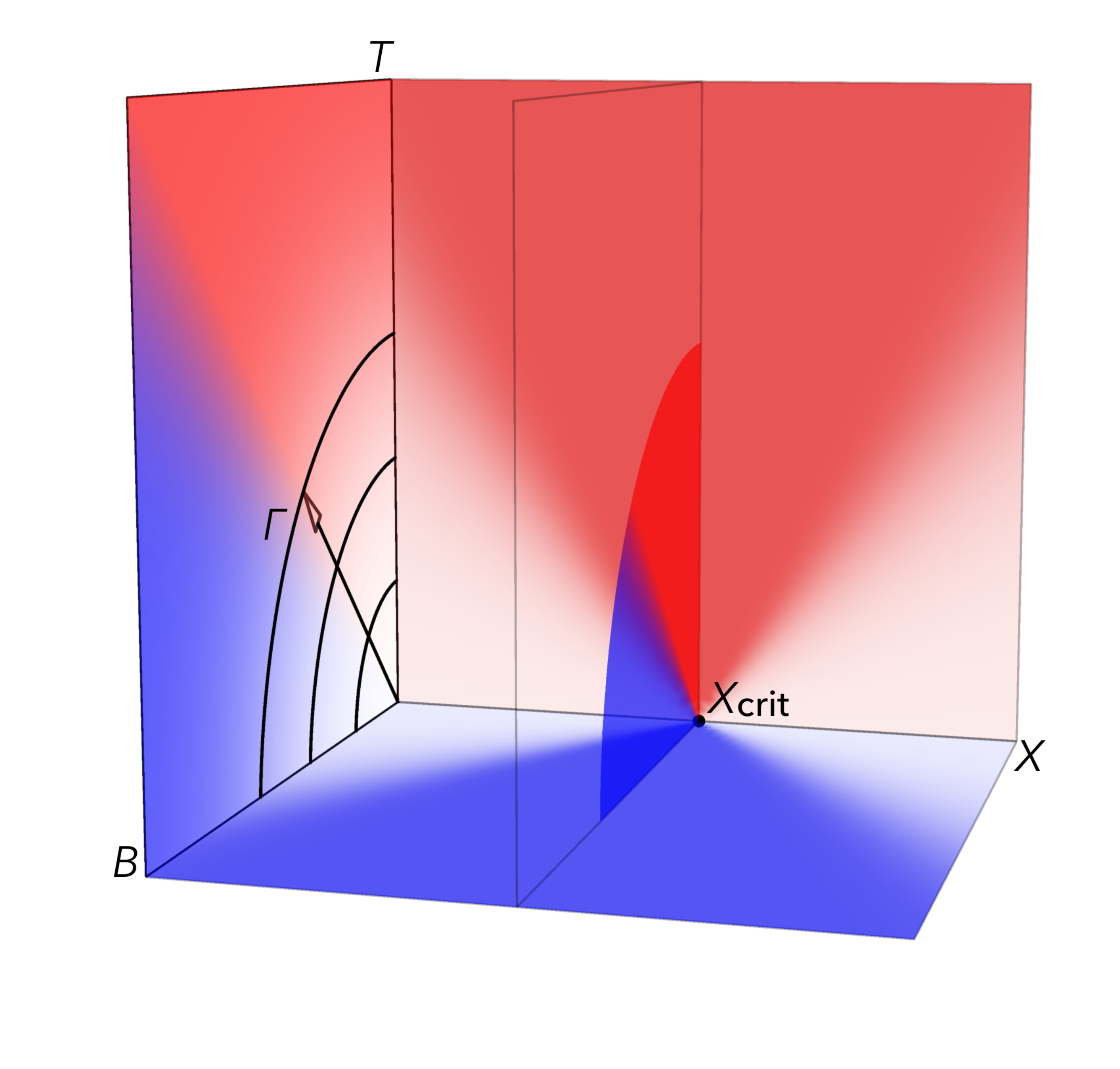} 
\caption{{\bf A quantum critical point where the scattering can be set by two orthogonal energy scales.} In Eq. \ref{eq:phenomenon} the magnetic field and temperature play identical roles in setting the energy scale of the scattering, such that they appear $T$-linear at small $B$ (red region), and $B$-linear at small $T=0$ (blue region). The data suggest that it is more useful to understand this physics as a balance of the two scales, which is phenomenologically captured by Eq. \ref{eq:phenomenon}. }
\label{fig:QCPBT} 
\end{figure}

High magnetic fields have played an important role in the study of QCPs, both as a tuning parameter to drive systems toward a QCP \cite{butch_quantum_2012,borzi_formation_2007,custers_break-up_2003} and in suppressing competing ordered states to reveal a QCP \cite{cooper_anomalous_2009}. Here the role of magnetic field is quite different; our measurements suggest that $B$-linear resistivity has the same origin as $T$-linear resistivity. This suggests a revision of the doping-temperature quantum critical phase diagram to include the role of an applied magnetic field, as shown in Fig. 5. Here the important role of the combined energy scale $\Gamma$ is immediately apparent: at comparable values of $B$ and $T$ the transport is linear in $\Gamma$ but not in $B$ or $T$ individually. Viewed in this way, magnetic fields provide another probe of the anomalous dynamics that lead to T-linear resistivity, with the added benefit that, unlike temperature, an applied magnetic field couples selectively to charge, spin and material anisotropies, limiting the possible theoretical descriptions of this phenomenon. It is striking that the scale factor $\eta/\alpha$ connecting magnetic field and temperature energy scales is close to unity and does not appear to change in the doping range considered, which is surprising because the quasiparticle effective mass changes considerably in the same doping range~\cite{shishido_evolution_2010}. This strongly suggests that the magnetoresistance in \basp\,originates from the direct effect of magnetic field on the quantum dynamics near its QCP, rather than the more conventional origin involving orbital motion about the Fermi surface. These measurements may point to a universal theoretical description of the physics of metals near a QCP where the scattering rate depends on the intricate interplay of energy scales \cite{sachdev_universal_1992, davison_holographic_2014, varma_phenomenology_1989}. The magnetoresistance may therefore prove crucial in elucidating the quantum dynamics of other systems that are thought to have vanishing energy scales, including the cuprate superconductors and heavy fermion systems \cite{weickert_low-temperature_2006, cooper_anomalous_2009}.

\section{Acknowledgements}
This work was completed with the support of the Laboratory Directed Research and Development Program of Lawrence Berkeley National Laboratory under the US Department of Energy Contract No. DE-AC02-05CH11231. The work at the National High Magnetic Field Laboratory is supported via NSF/DMR 1157490. We would like to thank Qimiao Si, Philip Phillips, Piers Coleman, Dmitrii Maslov, Andrey Chubukov, Rafael Fernandes, Luis Taillefer, Phuan Ong, Joe Orenstein, Andy Mackenzie, Ashvin Vishwanath, D. H. Lee, Joel Moore and P. W. Anderson for useful discussions.

%%%%%%%%%%%%%%%%%%%%%%%%%%%%%%%%%%%%%%%%%%%%%%%%%%%%%%%%%%%%%%%%%%%%%%%%%%%%%%%%%%%%%%%%%%%%%%%%%%%%%%%%%%%%%%%%%%%%%%%%%%%%%%%%%%%%%%%%%%%%%%%%%%%%%%%%%%%%%%%%%%%%%%%%%%%%%%%%%%%%%%%%%%%%%%%%%%%%%%%%%%%%%%%%%%%%%%%%%%%%%%%%%%%%%%%%%%%%%%%%%%%%%%%%%%%%%%%%%%%%%%%%%%%%%%%%%%%%%%%%%%%%%%%%%%%%%%%%%%%%%%%%%%%%%%%%%%%%%%%%%%%%%%%%%%%%%%%%%%%%%%%%%%%%%%%%%%%%%%%%

%\bibliography{/Users/jamesanalitis/Documents/Paper/pnictide_master}

%\pagebreak
\section{Supplementary Information}

\subsection{Materials and Methods}
 
Single crystals of \basp\, were grown by a self-flux method described elsewhere \cite{analytis_enhanced_2010}. Magnetoresistance was measured by a standard four-probe method and magnetic fields up to 65T were accessed at the NHMFL Pulsed Field Facility, Los Alamos National Laboratory. Contact resistances of around $\sim$1$\Omega$ were achieved by sputtering Au and attaching gold wires with EpoTek H20E. The magnetic field was oriented parallel to the crystallographic $c$-axis and orthogonal to the current direction. The composition for these materials was previously determined using XPS. Samples used in this study were taken from the same or similar batches and found to have the anticipated $T_c$, which correlates well with composition $x$.

\subsection{Determination of $\alpha/\eta$}

\begin{figure*}[ht]
\includegraphics[width=18cm]{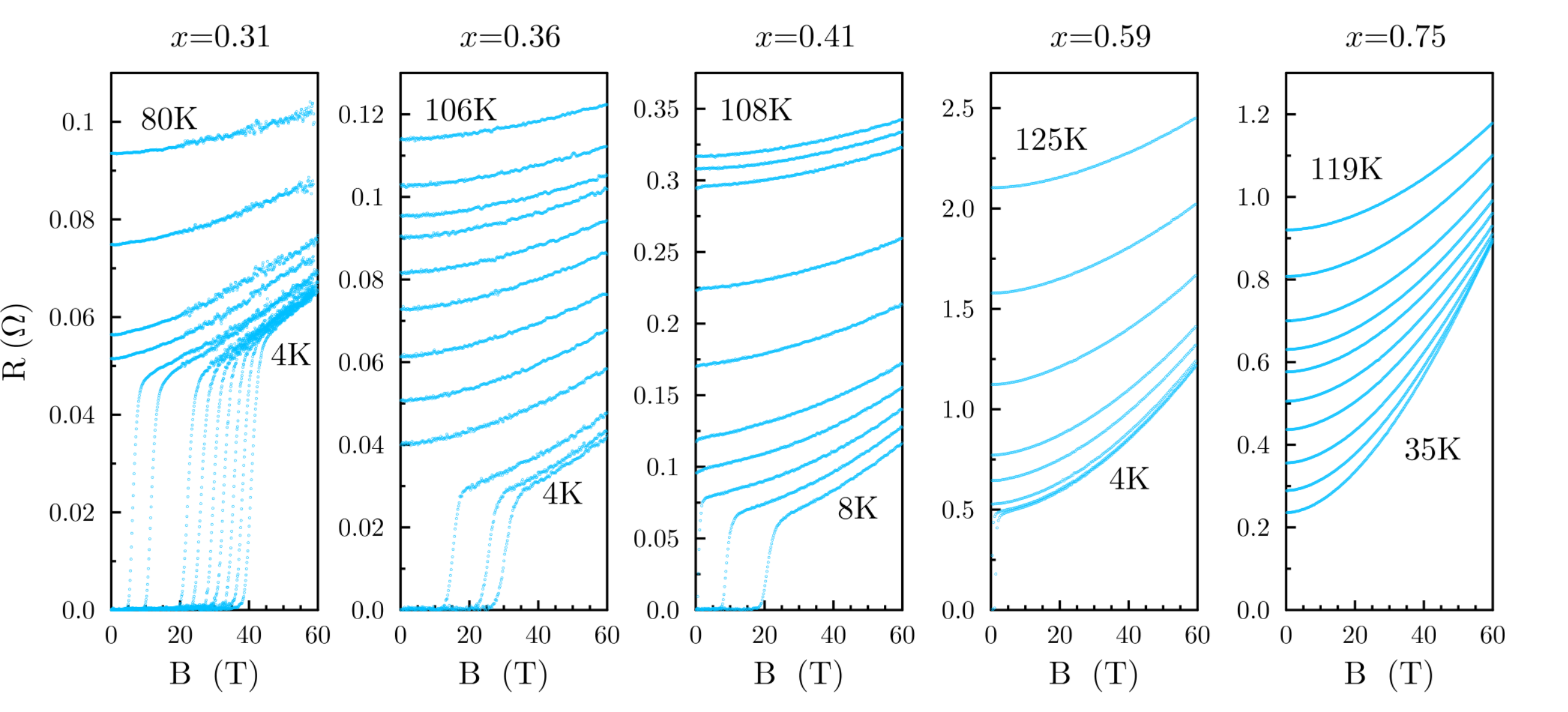} 
\caption{{\bf Resistance versus magnetic field for five different samples of \basp.} Each of the five panels in this figure shows the magnetoresistance at one of the dopings included in this study, starting with the doping nearest to the critical point on the left and going to far overdoped (nonsuperconducting) on the far right. Each curve was taken at a fixed temperature within the range specified in the figure.}
\label{fig:rawMR} 
\end{figure*}

\begin{figure*}[ht]
\includegraphics[width=18cm]{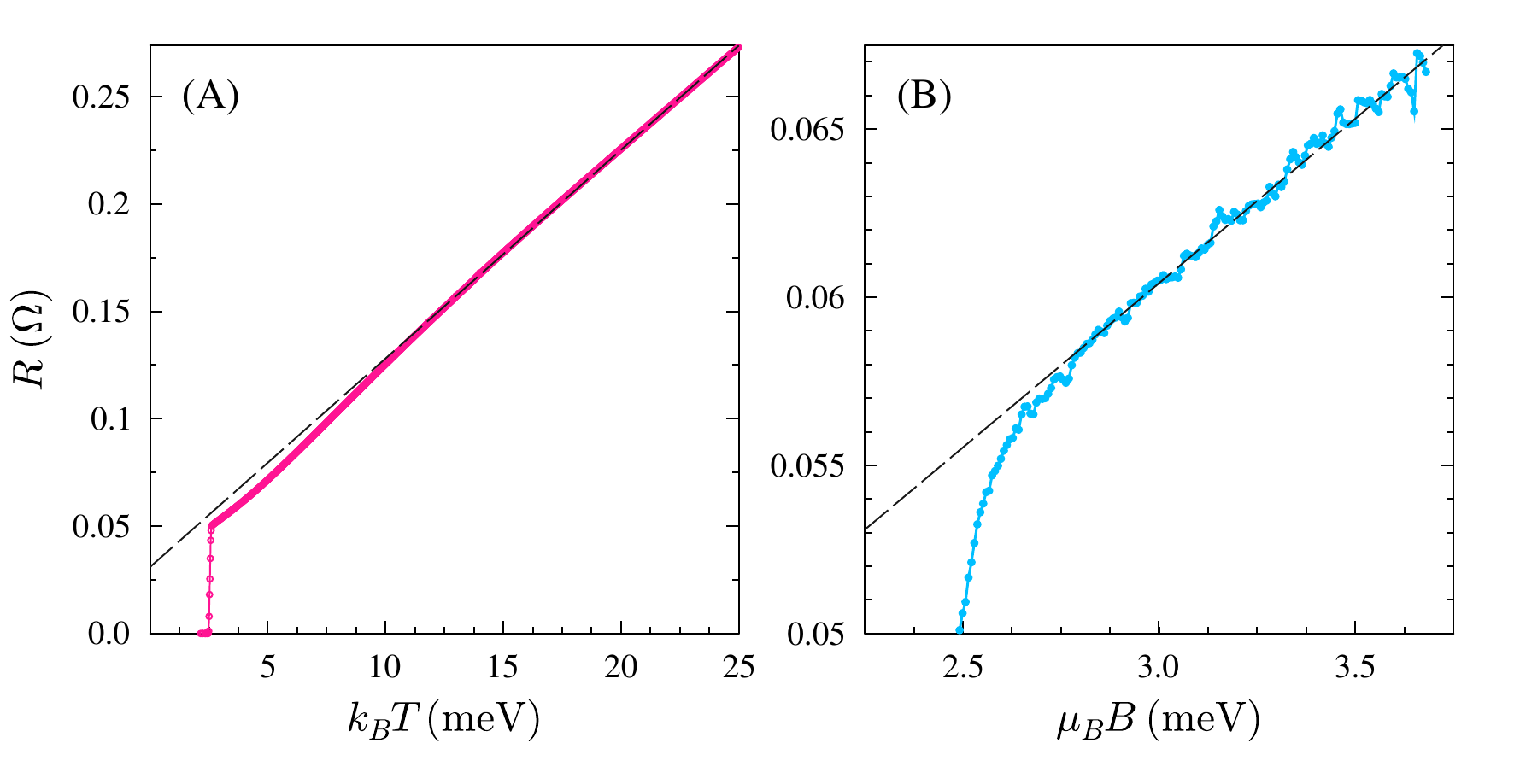} 
\caption{{\bf Fits to the linear portion of R(T) and R(B).} {\bf A} and {\bf B} show ressitance as a function of temperature and field, but with those two parameters converted to energy using $k_B$ and $\mu_B$ respectively. The fits to the linear regions give the same slope and intercept, suggesting $\eta/\alpha\approx 1$.}
\label{fig:Fits} 
\end{figure*}

Fig. \ref{fig:rawMR} shows the raw MR data as a function of field for different compositions, $x$. 
For $x = 0.31$ the low temperature MR curves are linear, and so the scale factor can be termined by the ratio of of the slopes of curves $R(T)$ and $R(B)$. The linear region is not sharply delimited in either case; the cross-over from $T$-squared to $T$-linear behavior is gradual and, similarly, the transition from superconducting to normal state as a function of $B$ is broad (perhaps due to superconducting fluctuations). Therefore conservative choices,  140K ($k_BT$=12.1meV) and 50T ($\mu_BB=$2.89meV), were made for the lower bounds of the linear portions of R(T) and R(B), respectively (see Fig. \ref{fig:Fits}). Dividing the slopes of the best-fit lines in these regions and normalizing to $\mu_B/k_B$ yields $\eta/\alpha = 1.01$. Taking the standard variation of the data about the best fit lines to constitute the error, the range of $\eta/\alpha$ that can be obtained from fit lines lying entirely inside the error bars is $1.01 \pm 0.07$. The range of residual normalized resistances, $\tilde{\rho}_0$, obtainable with those lines is $0.120 \pm 0.004$. 

\begin{figure*}[ht]
\includegraphics[width=18cm]{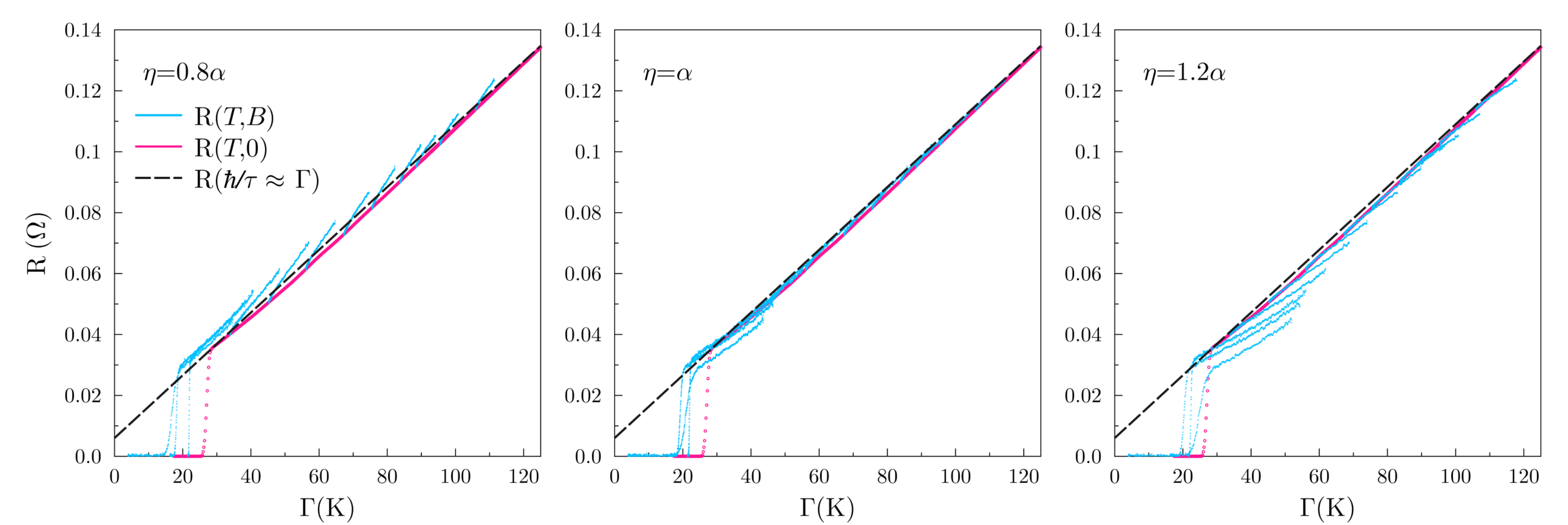} 
\caption{{\bf Resistance versus $\Gamma$ for different values of the scale factor.} Here we replot $R$ as a function of $\Gamma$ for the doping $x = 0.36$, but with the scale factor  increased (far right panel) or decreased (far left panel) from the value used in the main text. In each case the MR curves do not line up with the linear component of the zero field resistance curve (gray line), but either rise above or fall below it. The plot from Fig \ref{fig:rvg2} (A) with $\eta = \alpha$ is reproduced for comparison (center plot). Note that for any value of $\eta/\alpha$, the MR measured below $T_c$ is not co-linear with gray line, indicating the presence of a further energy scale (presumably associated with supercondcutivity), that is not captured by our ansatz, Eq. \ref{eq:phenomenon}.}
\label{fig:betavar} 
\end{figure*}

For samples with a higher phosphorous content, the lowest temperature MR curves show clear curvature, as expected for samples outside the critical region. Therefore, one cannot determine $\eta/\alpha$ for these samples as was done for $x=0.31$. However by replotting the data as a function of $\Gamma$ with different values for the scale factor, it is clear by eye that the high temperature data only fit to single, linear function of $\Gamma$ for $\eta/\alpha \approx 1$ (see Fig. \ref{fig:rvg2} of the main text). In Fig. \ref{fig:betavar} we show two slightly different choices of $\eta/\alpha$, 0.8 and 1.2. Even with this small variation, the deviation from the $\Gamma$-linear line is immediately apparent.

%\bibliography{qcp_master}

%\appendix

\end{document}